\documentstyle[12pt]{article}

\newcommand{\sect}[1]{\setcounter{equation}{0}\section{#1}}

\def\bseq{\begin{subequation}}  
\def\eseq{\end{subequation}}
\def\bsea{\begin{subeqnarray}}  
\def\esea{\end{subeqnarray}}

\def\Bar#1{\overline{#1}}                       

\def\Tilde#1{\widetilde{#1}}                    

\def\caja{\mathsurround=0pt}
\def\eqalign#1{\,\vcenter{\openup2\jot \caja
	\ialign{\strut \hfil$\displaystyle{##}$&$
	\displaystyle{{}##}$\hfil\crcr#1\crcr}}\,}

\def\Dot#1{{\kern0.5pt
     {#1} \kern-5.05pt \raise5.8pt\hbox{$\textstyle.$}\kern
0.5pt}}

\newcommand{\beq}{\begin{equation}}
\newcommand{\eeq}{\end{equation}}
\newcommand{\bea}{\begin{eqnarray}}
\newcommand{\eea}{\end{eqnarray}}
\newcommand{\ena}{\end{eqnarray}}

\renewcommand{\a}{\alpha}
\renewcommand{\b}{\beta}

\renewcommand{\d}{\delta}
\newcommand{\pa}{\partial}

\newcommand{\g}{\gamma}
\newcommand{\G}{\Gamma}

\newcommand{\e}{\epsilon}

\renewcommand{\l}{\lambda}
\renewcommand{\L}{\Lambda}
\newcommand{\m}{\mu}

\newcommand{\n}{\nu}

\newcommand{\p}{\pi}

\renewcommand{\o}{\omega}

\newcommand{\Phib}{\bar{\Phi}}

\def\Mb{\kern 2pt\mathchoice
	    {
	     \vbox{\hrule width10pt height 0.4pt depth 0pt
		 \kern 1.2pt\hbox{\kern -2pt$\displaystyle M$}}}
	    {
		 \vbox{\hrule width10pt height 0.4pt depth 0pt
		 \kern 1.2pt\hbox{\kern -2pt$\textstyle M$}}}
	    {
\vbox{\hrule width6pt height 0.4pt depth 0pt
		 \kern 1.0pt\hbox{\kern -2pt$\scriptstyle M$}}}
	    {
		 \vbox{\hrule width5pt height 0.4pt depth 0pt
		 \kern 0.8pt\hbox{\kern -2pt$\scriptscriptstyle M$}}}}

\def\Sb{\kern 2pt\mathchoice
	    {
		 \vbox{\hrule width6pt height 0.4pt depth 0pt
		 \kern 1.2pt\hbox{\kern -2pt$\displaystyle S$}}}
	    {
		 \vbox{\hrule width6pt height 0.4pt depth 0pt
		 \kern 1.2pt\hbox{\kern -2pt$\textstyle S$}}}
	    {
		 \vbox{\hrule width3.5pt height 0.4pt depth 0pt
		 \kern 1.0pt\hbox{\kern -2pt$\scriptstyle S$}}}
	    {
		 \vbox{\hrule width3pt height 0.4pt depth 0pt
		 \kern 0.8pt\hbox{\kern -2pt$\scriptscriptstyle S$}}}}

\def\Rb{\kern 2pt\mathchoice
	    {
		 \vbox{\hrule width5.5pt height 0.4pt depth 0pt
		 \kern 1.2pt\hbox{\kern -2.5pt$\displaystyle R$}}}
	    {
		 \vbox{\hrule width5.5pt height 0.4pt depth 0pt
		 \kern 1.2pt\hbox{\kern -2.5pt$\textstyle R$}}}
	    {
		 \vbox{\hrule width3.5pt height 0.4pt depth 0pt
		 \kern 1.0pt\hbox{\kern -2.2pt$\scriptstyle R$}}}
	    {
		 \vbox{\hrule width3pt height 0.4pt depth 0pt
		 \kern 0.8pt\hbox{\kern -2.2pt$\scriptscriptstyle R$}}}}

  \def\pp{{\mathchoice
	  {
	      \kern 1pt%
	      \raise 1pt
	      \vbox{\hrule width5pt height0.4pt depth0pt
		    \kern -2pt
		    \hbox{\kern 2.3pt
			  \vrule width0.4pt height6pt depth0pt
			  }
		    \kern -2pt
		    \hrule width5pt height0.4pt depth0pt}%
		    \kern 1pt
	   }
	    {
	      \kern 1pt%
	      \raise 1pt
	      \vbox{\hrule width4.3pt height0.4pt depth0pt
		    \kern -1.8pt
		    \hbox{\kern 1.95pt
			  \vrule width0.4pt height5.4pt depth0pt
			  }
		    \kern -1.8pt
		    \hrule width4.3pt height0.4pt depth0pt}%
		    \kern 1pt
	    }
	    {
	      \kern 0.5pt%
	      \raise 1pt
	      \vbox{\hrule width4.0pt height0.3pt depth0pt
		    \kern -1.9pt  
		    \hbox{\kern 1.85pt
			  \vrule width0.3pt height5.7pt depth0pt
			  }
		    \kern -1.9pt
		    \hrule width4.0pt height0.3pt depth0pt}%
		    \kern 0.5pt
	    }
	    {
	      \kern 0.5pt%
	      \raise 1pt
	      \vbox{\hrule width3.6pt height0.3pt depth0pt
		    \kern -1.5pt
		    \hbox{\kern 1.65pt
			  \vrule width0.3pt height4.5pt depth0pt
			  }
		    \kern -1.5pt
		    \hrule width3.6pt height0.3pt depth0pt}%
		    \kern 0.5pt
	    }
	}}

  \def\mm{{\mathchoice
		       {
			     \kern 1pt
	       \raise 1pt    \vbox{\hrule width5pt height0.4pt depth0pt
				  \kern 2pt
				  \hrule width5pt height0.4pt depth0pt}
			     \kern 1pt}
		       {
			    \kern 1pt
	       \raise 1pt \vbox{\hrule width4.3pt height0.4pt depth0pt
				  \kern 1.8pt
				  \hrule width4.3pt height0.4pt depth0pt}
			     \kern 1pt}
		       {
			    \kern 0.5pt
	       \raise 1pt
			    \vbox{\hrule width4.0pt height0.3pt depth0pt
				  \kern 1.9pt
				  \hrule width4.0pt height0.3pt depth0pt}
			    \kern 1pt}
		       {
			   \kern 0.5pt
	     \raise 1pt  \vbox{\hrule width3.6pt height0.3pt depth0pt
				  \kern 1.5pt
				  \hrule width3.6pt height0.3pt depth0pt}
			   \kern 0.5pt}
		       }}

\def\pd{{\kern0.5pt
		   + \kern-5.05pt \raise5.8pt\hbox{$\textstyle.$}\kern 
0.5pt}}

\def\pmd{{\kern0.5pt
		  \pm \kern-5.05pt
\raise6.3pt\hbox{$\textstyle.$}\kern1.5pt}}

\def\md{{\mathchoice
   {
      {{\kern 1pt - \kern-6.2pt \raise5pt\hbox{$\textstyle.$}\kern
1pt}}}
    {
      {{\kern 1pt - \kern-6.2pt \raise5pt\hbox{$\textstyle.$}\kern
1pt}}}
    {
      {\kern0.5pt - \kern-5.05pt
\raise3.4pt\hbox{$\textstyle.$}\kern0.5pt}}
    {
      {\kern0.5pt - \kern-5.05pt
\raise3.4pt\hbox{$\textstyle.$}\kern0.5pt}}}}

\newcommand{\ad}{{\dot{\alpha}}}

\newcommand{\Del}{\nabla}
\newcommand{\Delb}{\Bar{\nabla}}

\newcommand{\VEV}[1]{\left\langle #1 \right\rangle}

\begin{document}

\begin{titlepage}
{\hbox to\hsize{October 2000 \hfill
{Bicocca--FT--00--15}}}
{\hbox to\hsize{${~}$ \hfill
{BRX TH--481}}}
\begin{center}
\vglue .06in
{\Large\bf Noncommutative Supersymmetric \\
Gauge Anomaly }\\[.45in]
Marcus T. Grisaru\footnote{grisaru@brandeis.edu}
\footnote{Supported in 
part by National Science Foundation Grant 
PHY-00-70475, by INFN, MURST and the European Commission TMR program
ERBFMRX--CT96--0045, in which S.P. is associated to the University of 
Padova. }\\
{\it Department of Physics, Brandeis University, Waltham, MA 02454, USA}
\\
[.2in]
Silvia Penati\footnote{Silvia.Penati@mi.infn.it}\\
{\it Dipartimento di Fisica dell'Universit\`a di
Milano-Bicocca\\ 
and INFN, Sezione di Milano, piazza della Scienza 3, I-20126 Milano, 
Italy}\\[.8in]

{\bf ABSTRACT}\\[.0015in]
\end{center}

We extend the general method of \cite{us} to compute the consistent 
gauge anomaly for noncommutative 4d SSYM coupled to chiral matter. 
The choice of the minimal homotopy path allows us to obtain a 
simple and  compact result. We perform the reduction to components in the 
WZ gauge  proving that our result contains, as lowest component, the bosonic
chiral anomaly for noncommutative YM theories recently obtained in 
literature.  

${~~~}$ \newline

\end{titlepage}

\sect{Introduction}

In the recent literature a large number of papers have  appeared devoted to
various aspects of noncommutative field theory. Some of these papers have
concerned themselves with noncommutative supersymmetric field theories,
and a few have treated the subject in a superspace context \cite{NC}. In this
work, we address ourselves to the  superspace computation of the consistent
anomaly for NCSYM, using  some recently developed methods 
\cite{HOOS, OOSY, us}. We obtain  the noncommutative version of the
result in ref. \cite{us} and we show that the reduction to components
agrees with the consistent anomaly for the bosonic NCYM recently computed 
in ref. \cite{GBM}.

\sect{NC super Yang--Mills coupled to chiral matter}

The noncommutative $N=1$ supersymmetric Yang--Mills theory 
in four dimensions 
can be defined on a superspace described by bosonic 
noncommutative $x^{\mu}$ coordinates, $ [ x^{\m}, x^{\n} ] = i 
\Theta^{\m\n}$ and spinorial anticommuting cooordinates $ \{ \theta^\a ,
\theta^\b \} = \{ \bar{\theta}^{\dot\a} ,\bar{\theta}^{\dot\b} \} =0$
\cite{NC}.   
In chiral representation the constraints for the superspace covariant
derivatives are solved by (we use the obvious generalization of 
{\em Superspace} \cite{super} conventions) 
\beq
\Del_\a ~=~ (e^{-V})_{\star} \star D_\a (e^V)_{\star} \qquad \qquad
\Del_{\dot\a} ~=~ \bar{D}_{\dot\a} \nonumber\\
\qquad \qquad \Del_a ~=~ -i \{ \Del_\a , \Del_{\dot\a} \} 
\eeq
and the corresponding field strengths are given by
\bea  
&& W_\a ~=~ 
i \bar{D}^2 \left( ( e^{-V})_{\star} \star D_\a (e^V)_{\star} \right) 
~\equiv~ \bar{D}^2 \G_\a \nonumber \\ 
&& \Bar{W}_{\dot\a} ~=~ i D^2 \left( ( e^{V})_{\star} \star \bar{D}_{\dot\a} 
(e^{-V})_{\star} \right) ~\equiv~ D^2 \Bar{\G}_{\dot\a}
\label{connection}
\ena
where $\G_\a$ and $\Bar{\G}_{\dot\a}$ are the spinorial connections.
Moreover, we have defined $(e^V)_{\star} \equiv 1 + V + \frac{1}{2} V \star V 
+ \cdots$.

The dynamics of chiral scalar matter in the fundamental representation of
the gauge group minimally coupled to the gauge field 
is described by the action 
\beq
\int d^8z \, \bar{\Phi} \star (e^V)_{\star} \star \Phi
\label{action}
\eeq
where $\Phi$ and $\bar{\Phi}$ are chiral and antichiral superfields,
respectively ($\bar{D}_{\dot\a} \Phi = D_\a \bar{\Phi} =0$).
Here we have used the  notation $z \equiv (x, \theta, \bar{\theta})$.
This action  
is invariant under infinitesimal gauge transformations generated by 
(anti)chiral parameters $\L$ ($\bar{\L}$)
\bea
&& \d \phi ~=~ i \L \star \phi \qquad \quad 
\d \bar{\phi} ~=~ -i \bar{\phi} \star \bar{\L}  \noindent \\
&& \d (e^V)_{\star} ~=~ 
i \left( \bar{\L} \star (e^V)_{\star} - (e^V)_{\star} \star \L \right) \nonumber
\ena

In order to perform standard functional calculations we recall some basic
identities satisfied by the $\star$--product. Using the following 
property
\beq
\int d^4x \, (f \star g)(x) ~=~ \int d^4x \, f(x) \cdot g(x) 
\label{2}
\eeq
the functional derivative in the noncommutative case can be defined 
as usual
\bea
&& \frac{\d}{\d \phi(y)} \int d^4x \, 
(\phi_1 \star \phi_2 \star \phi_3)(x)
~=~ \int d^4x \, \frac{\d \phi_1(x)}{\d \phi_1(y)} \cdot 
(\phi_2 \star \phi_3)(x) \nonumber \\
&& = ~ \int d^4x \, 
\d^{(4)}(x-y) \, (\phi_2 \star \phi_3)(x) =  (\phi_2 \star \phi_3)(y)
\label{der}
\ena
The same definition holds for the functional derivative with 
respect to $\phi_2$ or $\phi_3$ since, by the cyclicity of the 
$\star$~--product, the field with respect to which we differentiate can be 
written as the first entry of the product. 
As a consequence of the previous results we have for instance
\beq
\int d^4y \d \phi_1(y) \frac{\d}{\d \phi_1(y)} 
\int d^4x \, (\phi_1 \star \phi_2 \star \phi_3)(x) 
~= ~ \int d^4y \, (\d \phi_1 \star \phi_2 \star \phi_3)(y)
\eeq
and more generally
\beq
\d \int d^4x \, \phi_1 \star \phi_2 \star \phi_3 
~=~ \int d^4x \left( \d \phi_1 \star \phi_2 \star \phi_3  ~+~
\phi_1 \star \d \phi_2 \star \phi_3 ~+~
\phi_1 \star \phi_2 \star \d \phi_3 \right)
\label{3}
\eeq
The generalization of the previous identities to superspace is 
straightforward. In particular, the functional derivative with respect
to chiral and antichiral superfields is defined as usual
\beq
\frac{\d \Phi(z) }{\d \Phi(z')} ~=~ \bar{D}^2 \d^{(8)}(z-z')
\qquad \quad 
\frac{\d \Phib(z) }{\d \Phib(z')} ~=~ D^2 \d^{(8)}(z-z') 
\eeq

We now compute the covariant propagator 
for the chiral scalar described by the action (\ref{action}).
We write the functional integral
\beq
\VEV{\Phib(z)} = \int D \Phi D \Phib \exp[ - \int d^8w~ \Phib \star 
(e^V)_{\star} \star \Phi ] \cdot  \Phib (z)
\label{vev1}
\eeq
and make a change of variable $\Phib \rightarrow  \Phib + \d \Phib$
under which the functional integral doesn't change. One gets then
\bea
&&0~=~\int D \Phi D \Phib \exp[ - \int d^8w~ (\Phib \star (e^V)_{\star} 
\star \Phi) (w) ] \nonumber \\
&&~~~~~~~~~~~~~~~~~~~
 \times [-\int d^8u~ (\d \Phib  \star (e^V)_{\star} \star \Phi)(u) 
\cdot \Phib (z) + \d \Phib (z) ]
\label{vev2}
\ena
Now taking the functional derivative $\d / \d \Phib (z') $ and
reinterpreting the
functional integral as giving an expectation value, we obtain
\beq
\VEV { \int d^8u D^2 \d^{(8)} (u-z')  \star ((e^V)_{\star} 
\star \Phi)(u) \cdot \Phib (z)- 
D^2 \d^{(8)} (z-z')  }  ~=~ 0 
\eeq
or
\beq
\VEV { \int d^8u \d^{(8)} (u-z')  \star  D^2 \Big[ ((e^V)_{\star} 
\star \Phi)(u) \cdot \Phib (z) \Big] } - 
D^2 \d^{(8)} (z-z')    ~=~ 0 
\eeq
Using (\ref{der}) we can write the previous equation as
\beq
D^2 (e^V)_{\star} \star \VEV{\Phi(z') ~\Phib (z)} 
~=~ D^2 \d^{(8)} (z'-z) 
\eeq
where the $\star$--product on the l.h.s. is with respect to the $z'$ variable.
One now proceeds as in the commutative case: 
$\star$--multiplying by $(e^{-V})_{\star}$ on the left we have
\beq
\Del^2 \star \VEV{\Phi(z') ~\Phib (z)} ~=~  (e^{-V})_{\star} 
\star D^2  \d^{(8)} (z'-z) ~=~
\Del^2 \star (e^{-V})_{\star} \star \d^{(8)} (z' - z) 
\eeq
where $\Del^2 \equiv \frac12 \Del^\a \star \Del_\a$. 
We proceed then multiplying by $\Bar{\Del}^2$ 
(remember that $\Delb^2 = \bar{D}^2$, so the product is the standard 
product) and extending 
$\Delb^2 \Del^2$ acting on the chiral superfield to the invertible operator
\beq
\Box_+ ~\equiv~ 
\Del^2 \Bar{\Del}^2 ~+~ \Bar{\Del}^2 \Del^2 ~-~ \Bar{\Del}^{\dot\a} 
\Del^2 \Bar{\Del}_{\dot\a} ~=~ \Box ~-~ i W^{\a} \star \Del_{\a} ~-~
\frac{i}{2} (\Del^{\a} \star W_{\a})  
\label{boxscalar4d}
\eeq
where
\beq
\Box \equiv \frac12 \Del^{\a \dot\a} \star \Del_{\a \dot\a} 
\eeq
We obtain the covariant scalar propagator (we interchange $z$ and
$z'$)
\beq
\VEV{\Phi(z) ~\Phib (z')} = \Bar{\Del}^2 \frac{1}{\Box_+} 
\star \Del^2 \star (e^{-V})_{\star} \star \d^{(8)} (z-z') 
\label{prop}
\eeq
where we have defined the inverse $1/\Box_+$ as the operator such that
$1/\Box_+ \star \Box_+ \star f = \Box_+ \star 1/\Box_+ \star f = f$, for any 
$f$. Moreover, we have used the identity $ \Bar{\Del}^2 \Box_+ = \Box_+
\Bar{\Del}^2$.

The short--distance behaviour of the propagator can be covariantly 
regularized \cite{HOOS}  by introducing an UV  
cut--off $M$ and multiplying it by $(\exp{(\Box_+/M^2)})_{\star}$ 
\beq
\VEV{\Phi(z) ~\Phib (z')}_{\rm reg} = (\exp{(\Box_+/M^2)})_{\star} \star
\frac{1}{\Box_+} 
\star \Bar{\Del}^2 \Del^2 \star (e^{-V})_{\star} \star \d^{(8)} (z-z') 
\label{propreg}
\eeq
Now, using the obvious identities
\beq
d (e^{\Box_+ t})_{\star} ~=~ dt \Box_+ \star ( e^{\Box_+ t})_{\star}
~=~ ( e^{\Box_+ t})_{\star} \star \Box_+ dt
\eeq
we have
\beq
(e^{\Box_+/M^2} )_{\star} ~=~ - \int_{\frac{1}{M^2}}^{\infty}
d (e^{\Box_+ t})_{\star} ~=~ - \int_{\frac{1}{M^2}}^{\infty} dt 
(e^{\Box_+ t})_{\star} \star \Box_+
\eeq
Multiplying on the right by $1/\Box_+$ we finally obtain
\beq
(e^{\Box_+/M^2} )_{\star} \star \frac{1}{\Box_+} 
~=~ - \int_{\frac{1}{M^2}}^{\infty} dt (e^{\Box_+ t})_{\star} 
\eeq
An alternative expression for the regularized propagator is then
\beq
\VEV{\Phi(z) ~\Phib (z')}_{\rm reg} = 
 - \int_{\frac{1}{M^2}}^{\infty} dt (e^{\Box_+ t})_{\star} 
\star \Bar{\Del}^2 \Del^2 \star (e^{-V})_{\star} \star \d^{(8)} (z-z') 
\label{propreg2}
\eeq

\sect{The noncommutative consistent anomaly} 

To compute the consistent anomaly due to chiral matter coupled to
supersymmetric noncommutative Yang--Mills, we follow the general procedure
used in Refs. \cite{OOSY,us} supplemented by the choice of a minimal 
homotopic path, as suggested in \cite{us2}.

We introduce a homotopic path $g(y,V)$, $y \in [0,1]$, satisfying the
boundary conditions $g(0,V) = 1$ and $g(1,V) = (e^V)_{\star}$ for any $V$. 
Moreover,
we define the inverse path $g^{-1}$ as given by $ g^{-1} \star g =
g \star g^{-1} = 1$.  
The extended superspace covariant derivatives and the
corresponding spinorial field strengths 
\bea
&& \Del_\a ~=~ g^{-1} \star D_\a g \qquad \qquad
\Del_{\dot\a} ~=~ \bar{D}_{\dot\a} 
\qquad \qquad \Del_a ~=~ -i \{ \Del_\a , \Del_{\dot\a} \} \nonumber \\
&& {\cal W}_\a ~=~ i \bar{D}^2 ( g^{-1}\star D_\a g) 
\qquad \, \, 
\Bar{\cal W}_{\dot\a} ~=~ i D^2 (g \star \bar{D}_{\dot\a} g^{-1}) 
\label{extder}
\ena 
satisfy the usual supersymmetry algebra.  
The homotopically extended action is given by 
\beq
S ~=~ \int d^8z ~\Phib \star g \star \Phi
\label{extendedaction}
\eeq
We write the effective action as 
\beq
{\mit\Gamma}[V] ~=~ {\mit \G}_{y=0}[V] + \int_0^1 dy \, \pa_y {\mit\G} ~=~
\int_0^1 dy \int d^8z ~\pa_y g(z)_{ij} \star \VEV{ \frac{\d S}{\d g(z)_{ij}}}
\eeq
or equivalently as
\beq
{\mit\Gamma}[V] ~=~ \int_0^1 dy \int d^8z ~\pa_y g(z)_{ij}  
\VEV{ \frac{\d S}{\d g(z)_{ij}}} 
\label{effact}
\eeq
where we have set ${\mit\G}_{y=0}[V] = 0$. 
Using the explicit expression (\ref{extendedaction}) we can formally 
write
\beq
\VEV{ \frac{\d S}{\d g(z)_{ij}}} ~=~ \VEV{(\Phi \star 
\Phib)(z)}_{ij} 
~\equiv~ \lim_{z \to z'} e^{\frac{i}{2}\Theta \pa_x \pa_{x'}}
\VEV{ \Phi(z) ~\Phib(z') }_{ij}
\label{regular}
\eeq
where $\Theta \pa_x \pa_{x'} \equiv \Theta^{\m \n}  \frac{\pa}{\pa x^\m}
\frac{\pa}{\pa {x'}^\n }$. On the r.h.s. the extended
propagator is present. Its expression is simply given by (\ref{prop}) where
all occurences of $(e^V)_{\star}$ have been replaced by $g$. In particular,
the covariant derivatives and the field strengths are the ones defined in
(\ref{extder}).  

Using the  regularizations 
(\ref{propreg}) or (\ref{propreg2}), 
the regularized expression for (\ref{regular}) is then given by
\bea
\VEV{ \frac{\d S}{\d g(z)}} &&=~
\lim_{z' \to z} e^{\frac{i}{2} \Theta \pa_x \pa_{x'}}
(\exp{(\Box_+/M^2)})_{\star} \star \frac{1}{\Box_+} 
\star \Bar{\Del}^2 \Del^2 \star (e^{-V})_{\star} \star \d^{(8)} (z-z') 
\nonumber \\
&&=~ -   \lim_{z' \to z} e^{\frac{i}{2} \Theta \pa_x \pa_{x'}} 
\int_{\frac{1}{M^2}}^{\infty} dt (e^{\Box_+ t})_{\star} 
\star \Bar{\Del}^2 \Del^2 \star (e^{-V})_{\star} \star \d^{(8)} (z-z') 
\nonumber \\
&&~~~~~~
\ena

To construct the consistent anomaly we consider the variation of ${\mit\G}$ 
due to an infinitesimal gauge transformation on $V$. Due to the identities 
(\ref{2}, \ref{3}), the formal derivation 
follows exactly the same steps as in the standard commutative case 
\cite{us}. Varying (\ref{regular}), after an integration by parts on $y$ 
we can write the consistent anomaly as
\bea
\d {\mit\G}[V] && \equiv ~ L ~-~ \frac{1}{16\pi^2} \int_0^1 \, dy \, X(y)
\nonumber \\
&& =~ \int d^8z \, \d (e^V)_{\star} \VEV{ \frac{\d S}{\d (e^V)_{\star}}}
~+~ \int_0^1 dy \int d^8z \int d^8z'' \, \d g(z'')_{kl} \pa_y g(z)_{ij} 
\nonumber \\
&&~~~~~~~\times \left[ \frac{\d}{\d g(z'')_{kl}} 
\VEV{ \frac{\d S}{\d g(z)_{ij}}}  
 ~-~ \frac{\d}{\d g(z)_{ij}} \VEV{ \frac{\d S}{\d g(z'')_{kl}}} \right]
\label{anomaly1}
\ena
Note that, due to the identity (\ref{2}), this expression can be
equally well written with $\star$--products replacing the usual products. 
Being the variation of an effective action, the anomaly (\ref{anomaly1}) 
automatically satisfies the WZ consistency condition.

We first concentrate on the explicit evaluation of the covariant term $L$.
From the identity (\ref{propreg}) we can write (we make repeated use of 
identity (\ref{2})) 
\bea
L &=& \int d^8z \lim_{z' \to z} \, \d (e^{V})_{\star} \star 
e^{\frac{i}{2} \Theta \pa_x \pa_{x'}}
(e^{\Box_+/M^2})_{\star} 
\star \frac{1}{\Box_+} \star \Bar{\Del}^2 \Del^2 \star (e^{-V})_{\star} 
\star \d^{(8)}(z-z') \nonumber\\
&=&  \int d^8z \lim_{z' \to z} \, (e^{-V})_{\star} \star \d (e^V)_{\star} 
\star e^{\frac{i}{2} \Theta \pa_x \pa_{x'}} (e^{\Box_+/M^2})_{\star}
\star \frac{1}{\Box_+} \star \Bar{\Del}^2 \Del^2  \d^{(8)}(z-z') \nonumber \\
&&~~~~~~~~~~~~ 
\label{cov1}
\eea
(a trace in group theory labels is understood)
where we have moved the factor of $(e^{-V})_{\star} $ in front since 
we are dealing with a trace, both in group theory  and 
superspace, and we have used the cyclicity property of the $\star$--product.  
{}From the identity
\beq
(e^{-V})_{\star} \star \d (e^V)_{\star} ~=~ 
- i\left(  \Lambda - ( e^{-V})_{\star} \star \bar{\Lambda} \star 
(e^V)_{\star} \right) ~\equiv~
- i (\Lambda - \tilde{\Lambda})
\eeq
it is easy to prove that the expression for the covariant term 
splits into the sum of a holomorphic and an anti-holomorphic contribution.
We can then concentrate only on the holomorphic part,  the 
$\tilde{\Lambda}$ term  being simply obtained by hermitean conjugation. 

{}Proceeding as in the commutative case,  we 
replace in (\ref{cov1})  $(e^{-V})_{\star} \star \d (e^V)_{\star} $ 
with $-i\L$, and first pull out a 
$\Bar{\Del}^2$ from the superspace measure which, due to the chirality 
property of the various quantities 
in the integrand, can only act on the 
$z'$  appearing in the $\d$-function (recall that we are first taking 
the limit $z' \to z$). Using $\Bar{\Del}^2 \Del^2 
\d^{(8)}(z-z') 
 \stackrel{\longleftarrow}{\Bar{\Del}^2}=\Bar{\Del}^2 \Del^2 
\Bar{\Del}^2 \d^{(8)}(z-z')= \Box_+ \Bar{\Del}^2\d^{(8)}(z-z')$ we obtain
\beq
L= -i\int d^6 z \lim_{z' \to z}\Lambda \star 
e^{\frac{i}{2} \Theta \pa_x \pa_{x'}}(e^{\Box_+/M^2})_{\star}
\Bar{\Del}^2  \d^{(8)}(z-z')  
\eeq
Now, we write 
\beq
\d^{(8)}(z-z') = \frac{M^4}{(2 \pi)^4} \int d^4k e^{iMk(x-x')}
\d^{(4)}(\theta - \theta')
\label{fourier}
\eeq
and move the $\exp(iMkx)$ factor past the  $(e^{\Box_+/M^2})_{\star}$ operator.
Neglecting terms in the exponential which eventually
do not contribute to the anomaly, we obtain
\bea
L &=&-i \frac{M^4}{(2 \pi)^4} \int d^4x d^2 \theta \int d^4k \, 
\Lambda \star 
\lim_{x' \to x} e^{\frac{i}{2} \Theta \pa_x \pa_{x'}} e^{iMkx} 
e^{-iMkx'} \\
&&\lim_{\theta' \to \theta} (e^{-k^2 +i k^a \Del_a/M 
+\Box/M^2 -i W^\alpha \star \Del_\alpha/M^2 -i(\Del^{\alpha} \star 
W_{\alpha})/(2M^2)})_{\star} \Bar{\Del}^2 \d^{(4)} (\theta - \theta') 
\nonumber 
\eea
Now, we observe that 
\beq
\lim_{x' \to x} e^{\frac{i}{2} \Theta \pa_x \pa_{x'}} 
e^{iMkx} e^{-iMkx'} ~=~ e^{iMkx} \star e^{-iMkx} ~=~ 1
\label{one}
\eeq
Moreover, performing the limit on the spinorial coordinates we obtain 
a zero result except from terms, in the
expansion of the exponential, that can produce a factor of $\Del^2$ 
which together with the $\Bar{\Del}^2$ remove the $\d^{(4)}(\theta -
\theta')$. The only non--vanishing contribution comes from the second 
order term $1/2! (W^\a \star \Del_\a)^2$ which also has the correct  
$1/M^4$ factor to cancel the overall $M^4$. Therefore, one 
can now  take the limit $\theta' \rightarrow \theta$, remove the regulator,
 $M \to \infty$, perform the
$k$-integration of the remaining $e^{-k^2}$ factor, and obtain the final
form of the covariant anomaly
\beq
L= -\frac{i}{8 \pi^2} \int d^6 z ~{\rm Tr} \Big[ \, \Lambda \star W^\alpha
\star W_\alpha \, \Big] ~+~ {\rm {h.\,c.}} 
\label{covan}
\eeq

\vskip 15pt
We now focus on the evaluation of the consistent term in (\ref{anomaly1}). 
We consider the first contribution in that expression (the
second one is simply obtained by interchanging $\d g$ with $\pa_yg$)
\bea
&& \int_0^1 dy \, \int d^8z \, \int d^8z'' \d g(z'')_{kl} \pa_y g(z)_{ij}
\frac{\d}{\d g(z'')_{kl}} \VEV{\frac{\d S_y}{\d g(z)_{ij}}} \nonumber \\
&& =~ - \int_0^1 dy \, \int d^8z \pa_y g(z) \,
\lim_{z' \to z}  e^{\frac{i}{2} \Theta \pa_x \pa_{x'}} \,
\d \int_{\frac{1}{M^2}}^{\infty} dt \,
(e^{\Box_+ t})_{\star} \star \Delb^2 \Del^2 \star g^{-1} \star 
\d^{(8)}(z-z') \nonumber \\
&&~~~~~~~~~~~~~~~~~~~~~~~~
\label{first} 
\ena
The calculation follows exactly the one for the commutative
case, so that we list here only the main steps, referring to \cite{us} for
more details. 

Starting from (\ref{first}) we first perform the variation of 
the single operators. Using the identities 
\bea
\delta g^{-1} &=& - g^{-1} \star \delta g \star g^{-1} \nonumber \\
\delta \Del_\alpha &=& \delta (g^{-1} \star D_\alpha g)=
[\Del_\alpha ,g^{-1} \star \d g]_{\star} \nonumber \\
\d \Del^2 \star g^{-1} &=& - g^{-1} \star \d g \star \Del^2 \star 
g^{-1}  \nonumber\\
\delta (e^{\Box_+ t})_{\star} &=& \int_0^t ds (e^{\Box_+s})_{\star} 
\star \delta \Box_+ \star (e^{\Box_+(t-s)})_{\star} \nonumber \\
\d \Box_+ \Big|_{on~chirals} &=& \d \Delb^2 \Del^2 ~=~
\Delb^2 [ \Del^2, g^{-1} \star \d g]_{\star}
\eea
we obtain
\bea
&& \int_0^1 dy \int d^8z  \, \pa_y g(z) \lim_{z' \to z}
e^{\frac{i}{2} \Theta \pa_x \pa_{x'}} \,
\left[\, \int_{\frac{1}{M^2}}^\infty 
dt \,(e^{\Box_+t})_{\star} \star \Bar{\Del}^2 g^{-1} \star \d g \star 
\Del^2 \star g^{-1} \right. \nonumber \\
&&\left. - \int_{\frac{1}{M^2}}^\infty  dt \int_0^t ds
(e^{\Box_+s})_{\star} \star \Bar{\Del}^2 [ \Del^2 , g^{-1} \star 
\d g]_{\star} \star (e^{\Box_+(t-s)})_{\star} \star \Delb^2
\Del^2 \star g^{-1} \right] \star \d^{(8)}(z-z') \nonumber \\
&&~~~~~~~~~~ 
\label{varia}
\eea
Here $[A,B]_{\star}$ is the Moyal bracket $[A,B]_{\star} = A \star B
- B \star A$. Now, expanding the commutator in the second line 
and using the cyclicity properties of the 
trace and the $\star$--product, it is easy to see that the first term
eventually cancels when added to the second contribution in 
(\ref{anomaly1}) (the one obtained by interchanging $\d g$ with $\pa_yg$). 
The second term from the commutator
gives instead a nontrivial contribution which we rewrite in the 
following way
\bea
&&\int_{\frac{1}{M^2}}^\infty  dt \int_0^t ds
(e^{\Box_+s})_{\star} \star \Delb^2  g^{-1} \star \d g \star \Del^2 
\star (e^{\Box_+(t-s)})_{\star} \star \Delb^2
\Del^2 \star g^{-1} {~~~~~~~~~~~~~~~~~~~~} \nonumber\\
&&{~~~~~~~~~~~~}=\int_{\frac{1}{M^2}}^\infty  dt \int_0^t ds
(e^{\Box_+s})_{\star} \star \Delb^2  g^{-1} \star \d g \star \Del^2 
\star \Delb^2 (e^{\Del^2
\Delb^2(t-s)})_{\star} \star \Del^2 \star g^{-1}  \nonumber\\
&&{~~~~~~~~~~~~}= \int_{\frac{1}{M^2}}^\infty  dt \int_0^t ds
(e^{\Box_+s})_{\star} \star \Delb^2  g^{-1} \star \d g \star 
\frac{\pa}{\pa t} (e^{\Box_- (t-s)})_{\star} \star \Del^2 \star g^{-1} 
\label{consistent1}
\eea
where we have defined
\beq
\Box_- ~=~ \Del^2 \Delb^2 + \Delb^2 \Del^2 - \Del^\a \star \Delb^2 \Del_\a
~=~ \Box ~-~ i \Tilde{\cal W}^{\dot\a} \star \Delb_{\dot\a} 
~-~ \frac{i}{2} ( \Delb^{\dot\a} \Tilde{\cal W}_{\dot\a})
\eeq
with 
$\Tilde{\cal W}_{\dot\a} = -g^{-1} \star \Bar{\cal W}_{\dot\a} 
\star g$. To obtain the expression (\ref{consistent1}) we have used
the identity
\beq
\Del^2 \Delb^2 \star (e^{\Del^2 \Delb^2 (t-s)})_{\star} \star
~=~ \frac{\pa}{\pa t} (e^{\Del^2 \Delb^2 (t-s)})_{\star} \star
\eeq
which can be easily proved by Taylor expanding the exponential. 
 
The integration by parts of the $\pa_t$ derivative produces an
integrated term which cancels the first term in (\ref{varia}) 
and a second term which reads
\bea
&& - \lim_{z' \to z}  e^{\frac{i}{2} \Theta \pa_x \pa_{x'}} \,
\int_0^1 dy \int d^8z  \int_0^{\frac{1}{M^2}} ds \\
&&~~~~~~\pa_y g(z) \star
(e^{\Box_+s})_{\star} \star
\Delb^2 g^{-1} \star \d g \star (e^{\Box_-(1/M^2 -s)})_{\star} 
\star \Del^2 \star g^{-1} \star \d^{(8)} (z-z') \nonumber
\ena
This expression can be
manipulated in the same manner as we treated the covariant anomaly $L$.
We introduce a momentum basis for the $\d^{(4)}(x-x')$ factor (see
eq. (\ref{fourier})) and 
let the exponentials act on the $e^{ikx}$ factor.
This operation allows us to bring the $e^{ik(x-x')}$ term in front,
while producing factors of $Mk$ in the various exponentials.
The $x' \to x$ limit can be performed now using the identity (\ref{one}). 
In the limit $M^2 \to \infty$ and $\theta' \to \theta$ 
factors from the expansion of the exponentials proportional 
to $({\cal W}^\a \star\Del_\a)/M^2$, $(\Del^{\a} \star {\cal W}_{\a})/M^2$,
$(\Tilde{\cal W}^\ad \star \Delb_\ad) /M^2$ and $(\Delb^{\ad} 
\Tilde{\cal W}_{\ad})/M^2$ 
give the relevant contributions, cancelling the overall $M^2$ and
the $\d^{(4)}(\theta -\theta')$ factor. 
Adding the contribution obtained by $\d g \leftrightarrow \pa_yg$, 
the final result takes the form 
\bea
X(y) &=& 
   2i \, \int_0^1 dy\int d^8z\,
   {\rm Tr} \, h_1 \star \left(\left[ {\cal D}^\alpha h_2,{\cal W_\alpha}
   \right]_{\star} +\left[\bar{\cal D}_{\dot\alpha}h_2,
   \widetilde{\cal W}^{\dot\alpha}\right]_{\star}
   +\left\{h_2,\bar{\cal D}_{\dot\alpha}
   \widetilde{\cal W}^{\dot\alpha}\right\}_{\star} \right)
\nonumber\\
   &=& -2i \, \int_0^1 dy \int d^8z\, {\rm Tr} \left(
   h_1 \star \left[\bar{\cal D}^{\dot\alpha}h_2,
   \widetilde{\cal W}_{\dot\alpha}\right]_{\star}
   ~+~h_2 \star \left[ {\cal D}^\alpha h_1,{\cal W_\alpha}\right]_{\star}
   \right) 
\nonumber \\
   &=& -\frac{2}{3}i \, \int_0^1 dy \int d^8z\, {\rm Tr_s} \left(
   h_1 \star (\bar{\cal D}^{\dot\a} h_2) \star \Tilde{\cal W}_{\dot\a}
   ~+~h_2 \star ({\cal D}^{\a} h_1) \star {\cal W}_{\a} \right)  
\label{xterm}
\eea
where $h_1\equiv 
g^{-1} \star \delta g$ and~$h_2\equiv g^{-1} \star \pa_yg$. Moreover 
we have defined ${\cal D}_{\a} A \equiv \{ \Del_\a , A ]_{\star}$ 
for any scalar or spinor object $A$ in the adjoint representation 
of the gauge group. In the last equality ${\rm Tr_s}$ is the symmetrized 
trace defined as
\bea
&& {\rm Tr}_s ( A \star B^\a \star C_\a) \equiv  {\rm Tr} \left[ A \star 
\Big( B^\a \star C_\a - C_\a \star B^\a \Big) \right. \\
&&~~~ +~\left. B^\a \star \Big( C_\a \star A + A \star C_\a \Big) 
~-~ C_\a \star \Big( A \star B^\a + B^\a \star A \Big) \right] \nonumber
\ena
for any scalar $A$ and spinors $B_\a$, $C_\a$.
Using the noncommutative extended Bianchi identities ${\cal D}^\a {\cal W}_\a
+ \bar{\cal D}^{\dot\a} \Tilde{\cal W}_{\dot\a} =0$, one can easily show
that the previous expression is antisymmetric under the exchange 
$h_1 \leftrightarrow h_2$.

As in the commutative case \cite{OOSY,us},  
two different choices of homotopic paths lead to two cohomologically 
equivalent expressions for the consistent anomaly.
Following Refs. \cite{us2, us}, we choose the 4D, noncommutative
$N$ = 1 supersymmetric Yang-Mills gauge theory ``minimal'' homotopy operator
\beq
g ~\equiv~ 1 ~+~ y\, (~ (e^V)_{\star} \, - \,1 ~) 
\label{minimal}
\eeq
The advantage of this choice is easily understood from the identities
\beq
\delta g ~=~ -iy\, \left( (e^V)_{\star} \star \Lambda-\bar{\Lambda} \star 
(e^V)_{\star}\right) \qquad \qquad
\pa_y g ~=~ \left( (e^V)_{\star} - 1 \right)
\eeq
which allow us to express $h_1$ and $h_2$ in (\ref{xterm}) still as functions
of $(e^V)_{\star}$. However, since in the covariant derivatives the inverse
$g^{-1}$ is present, the final expression for the noncommutative anomaly 
is necessarily a non--polynomial function of $(e^V)_{\star}$. This signals 
the presence of a no--go theorem analogue
to the one proved for the standard commutative SYM \cite{FGPS}. 

With the minimal choice for the homotopy the 4D, noncommutative
$N$ = 1 supersymmetric  Yang-Mills consistent anomaly is given by
the imaginary part  of a superaction, ${\cal A}_{\rm BGJ}(\L,\bar{\L}) 
= {\cal {I}{\rm m}} [ {\Tilde {\cal  A}}{}_{\rm BGJ}(\L) \,]$ where
\bea
&&\Tilde{\cal A}{}_{\rm BGJ}(\L) ~=~ 
\frac{1}{4\pi^2} \, \int \, d^8z \Big\{ \,
{\rm Tr} ( \L \G^\a \star W_\a ) \nonumber \\
&&~~~~~~~~~~~~~~~-~ \frac13 \, \int_0^1 \, dy \, y \, 
{\rm Tr}_s \Big( [(e^V)_{\star} 
\star {\cal G} \star \L] \star \pi^\a \star {\cal W}_\a \nonumber \\
&&~~~~~~~~~~~~~~~~~~~~~~~~~~~~+~ 
[ \, I ~-~ (e^V)_{\star} \star {\cal G} \, ] \star
[ \, \tilde{\pi}^{\dot \a} \star \L \, ] \star \Tilde{\cal W}_{\dot\a}\,  
\Big) \, \Big\} 
\label{BGJ3}
\ena
Here we have defined
\beq 
\eqalign{ {~~~~~~}
{\cal G} &\equiv~ \Big[ ~ 1 ~+~ y \, (\, (e^V)_{\star} \, - \,1  \,) ~ 
\Big]^{-1}  
~~~,~~~~ W_{\a} ~\equiv~ {\cal W}_{\a}(y = 1) ~=~ \bar{D}^2 \G_{\a} \cr
\p_{\a} &\equiv~  (e^V)_{\star} \star {\cal G} \star {\cal G} \star 
\G_{\a} {~~~~~~~~~~\,~~~} , {~\,~~~} \Tilde{\p}_{\dot \a} ~\equiv~e^V \, {
\cal G} \star \tilde{\G}_{\dot \a} \star {\cal G}  \cr  
{\cal W}_{\a} &\equiv~  {\bar D}{}^2 ( {\cal G} \star D_{\a} {\cal G
}^{-1} )  ~~~~~~~~~~~~~~~~,~~~ {\Tilde {\cal W}}{}_{\Dot \a} ~
\equiv~  {\cal G}\star [ D{}^2 ( {\cal G}^{-1} \star {\bar D}{}_{\dot \a}
{\cal G} ) \, ] \star {\cal G}^{-1} }
\label{new}
\eeq
Explicitly, the extended field strengths are given by
\bea
{\cal W}_\a &\equiv& y (e^V)_{\star} \star {\cal G} \star \o_\a \nonumber \\
\o_\a &\equiv& W_\a ~-~ (1 - y) \Big[ \tilde{\G}^{\dot\a} \star {\cal G} 
\star \G_a
~+~ (1 - y) \tilde{\G}^{\dot\a} \star {\cal G} \star \tilde{\G}_{\dot\a} 
\star {\cal G} \star \G_\a
\nonumber \\
&&~~~~~~- \frac{i}{2} ( \bar{D}^{\dot\a} \tilde{\G}_{\dot\a} ~-~ i 
\tilde{\G}^{\dot\a} \star \tilde{\G}^{\dot\a}) \star {\cal G} \star 
\G_\a \Big] \\
\Tilde{\cal W}_{\dot\a} &\equiv& 
y (e^V)_{\star} \star {\cal G} \star \tilde{\o}_{\dot\a} 
\nonumber \\
\tilde{\o}_{\dot\a} &\equiv& \Tilde{W}_{\dot\a} ~+~ (1-y) \Big[ 
\tilde{\G}_a \star {\cal G} \star \G^\a ~-~ (1-y) \tilde{\G}_{\dot\a} 
\star {\cal G} \star \G^\a \star {\cal G} \star \G_\a \nonumber \\
&&~~~~~- \frac{i}{2} ( D^\a \G_\a ~+~ i\G^\a \star \G_\a ) \Big]
\label{fextended}
\ena
All the extended functions appearing on the l.h.s. of 
(\ref{new}-- \ref{fextended}) are expressed
in terms of standard connections and field strengths of the NCSYM theory. 
The tilde quantities on the r.h.s. of these equations are defined as
$\tilde{A} \equiv (e^{-V})_{\star} \star A  \star (e^V)_{\star}$.
In the derivation of (\ref{BGJ3}--\ref{fextended}) we have
made repeated use of the identities
\beq
D_\a (e^V)_{\star} ~=~ -i (e^V)_{\star} \star \G_\a \qquad , \qquad 
\bar{D}_{\dot\a} (e^V)_{\star} ~=~ i (e^V)_{\star} \star
\Tilde{\G}_{\dot\a} 
\eeq
which follow from the definitions (\ref{connection}), and
\beq
{\cal G} \star (e^V)_{\star} ~=~ (e^V)_{\star} \star {\cal G} 
\eeq
\beq
y \, {\cal G} \star (e^V)_{\star} ~=~ 1 ~-~ (1 - y) \, {\cal G} \quad ,
\quad y \, ((e^V)_{\star} - 1) \star {\cal G} ~=~ 1 ~-~ {\cal G}
\eeq 
which are a consequence of the definition of the inverse function ${\cal G}$.

\sect{The physical bosonic component}

We perform the reduction to components of eq. (\ref{BGJ3}) and evaluate 
explicitly the physical bosonic term. This amounts to compute the superspace
integral and set fermions and auxiliary fields to
zero (auxiliaries can be neglected since they would produce higher order 
contributions).

Since the noncommutativity 
features of the theory do not affect the superfield structure, the definition
of the WZ gauge can be inherited from the commutative SYM theory 
\cite{super}. 
Therefore, we perform the reduction by choosing the  gauge 
\beq
V| ~=~ DV| ~=~ D^2 V| ~=~ 0
\eeq
where ``$|$'' means evaluation at $\theta = \bar{\theta} =0$. The basic 
ingredients required in the calculation are the physical bosonic components
of the connections and the field strengths. From the identities
\beq
D_{\a} \, \bar{D}_{\dot \a} \, V| ~=~ - \, A_{\a \dot \a}
\qquad \qquad 
\bar{D}_{\dot \a} \, D_{\a} \, V| ~=~ \, A_{\a \dot \a} 
\eeq
one easily obtains 
($\rightarrow$ indicates that fermions and auxiliary fields have been 
neglected)
\begin{itemize}
\item
Spinorial connections
\bea
&& \bar{D}_{\dot \a} \G_{\a}| ~=~ i A_{\a \dot \a} \qquad
D_{\b} \, \bar{D}^2 \, \G_{\a} | ~\rightarrow~ f_{\a\b} \nonumber \\
&& D_{\a} \, \bar{\G}_{\dot \a}| ~=~ i A_{\a \dot \a} \qquad
D^2 \, \bar{D}_{\dot \b} \, \bar{\G}_{\dot \a}| ~\rightarrow~ 
\bar{f}_{\dot \a \dot \b} ~+~ \pa_{~\dot \b}^{\b} \, A_{\b \dot \a}
\ena
\item
Vector connection
\bea
&& \G_{a} | ~=~ A_{\a \dot \a} \nonumber \\
&& D_{\b} \bar{D}_{\dot \b} \G_{a}| ~\rightarrow~ -i \, C_{\dot \a
\dot \b} \, f_{\a \b} \qquad
D^{\b} \, \bar{D}^2 \, D_{\b} \, \G_{a}| ~\rightarrow~ - 
\pa^{\b}_{~\dot\a} \, f_{\a \b}   \\
&& \bar{\G}_{a} | ~=~ A_{\a \dot \a} \nonumber \\
&& D_{\b} \bar{D}_{\dot \b} \bar{\G}_{a}| ~\rightarrow~ i \, C_{\a\b} 
\, \bar{f}_{\dot \a \dot \b} ~+~ i \, \pa_{\b \dot \b} \, A_{\a \dot \a}
\qquad
D^{\b} \, \bar{D}^2 \, D_{\b} \, \bar{\G}_{a}| ~\rightarrow~ - 
\pa^{~\dot \b}_{\a} \, \bar{f}_{\dot \a \dot \b} \nonumber
\ena
\item
Field strengths
\beq
D_{\b} \, W_{\a}| ~\rightarrow~ f_{\a \b} \qquad \qquad 
\bar{D}_{\dot \a} \, \Bar{W}_{\dot \b}| ~\rightarrow~ \bar{f}_{\dot \a
\dot \b} 
\eeq
\end{itemize}
Using these identities we can compute the relevant bosonic components of 
the quantities which enter eq. (\ref{BGJ3}). For $\omega_{\a}$ and 
$\tilde{\omega}_{\dot \a}$ we find
\bea
&& D_{\b} \, \omega_{\a}| ~\rightarrow~ f_{\a \b} ~-~ i(1-y) \, 
{A_{\b}}^{\dot \a} \star A_{\a \dot \a} \nonumber \\
&& D^2 \, \bar{D}_{\dot \b} \, \omega_{\a}| ~\rightarrow~ 
-(1-y) \, {A^{\b}} _{\dot \b}\star f_{\a\b} ~-~ i(1-y)^2 \, 
A_{\b \dot\b} \star A^{\b \dot\a} \star A_{\a \dot\a} \\
&& \bar{D}_{\dot\b} \, \tilde{\omega}_{\dot\a}| ~\rightarrow~ 
\bar{f}_{\dot\a \dot\b} ~-~ i(1-y) \, 
{A^{\a}}_{\dot\a} \star A_{\a \dot \b} \nonumber \\
&& D_{\b} \, \bar{D}^2 \, \tilde{\omega}_{\dot\a}| ~\rightarrow~ 
-(1-y) \,{ \bar{f}^{~\dot\b}} _{\dot\a}\star A_{\b \dot\b} 
~+~ iy(1-y) \, A_{\a \dot\a} \star A^{\a \dot\b} \star A_{\b \dot\b} 
\nonumber \\
&&~~~~~~~~~~-~ i(1-y) \, A_{\b \dot\b} 
\star A_{\a \dot\a} \star A^{\a \dot\b} ~+~ i{\pa_{\b}}^{\dot\b} \, 
\bar{f}_{\dot\a
\dot\b} ~+~ (1-y) \, {\pa_{\b}}^{\dot\b} \, ({A^{\a}} _{\dot\a}\star A_{\a 
\dot\b}) \nonumber
\ena
In the derivation of these components we made use of the Bianchi
identities 
\beq
{\pa^{\b}} _{\dot\b}\, A_{\b\dot\a} ~+~ \pa^{\b}_{~\dot\a} \,
A_{\b \dot\b} ~=~ -2 \bar{f}_{\dot\a\dot\b} ~+~ i \,
[ \, {A^{\b}}_{\dot\a} \, , \, A_{\b \dot\b} \, ]_{\star}
\eeq
Proceeding in the same manner, for $\pi_{\a}$ and $\tilde{\pi}_{\dot \a}$ 
we find
\bea
&&\bar{D}_{\dot\b} \, \pi_{\a}| ~\rightarrow~ i \, A_{\a \dot\b} \nonumber \\
&&D_{\b} \bar{D}^2 \, \pi_{\a}| ~\rightarrow~ f_{\a\b} ~-~ i(1-2y) \,
{A_{\b}}^{\dot\b} \star A_{\a\dot\b} \nonumber \\
&&D_{\b} \, \tilde{\pi}_{\dot\a}| ~\rightarrow~ 
i \, A_{\b \dot\a} \\
&&D^2 \bar{D}_{\dot\b} \, \tilde{\pi}_{\dot\a}| ~\rightarrow~ 
\bar{f}_{\dot\a\dot\b} ~+~{ \pa^{\b} }_{\dot\b}\, A_{\b \dot\a}
~+~ iy \, {A^{\b}}_{\dot\b} \star A_{\b\dot\a} ~-~ i(1-y) \, 
{A^{\b}}_{\dot\a} \star A_{\b \dot\b} \nonumber  
\ena
We are now ready to compute the physical bosonic components contained
in the supersymmetric anomaly (\ref{BGJ3}). In that equation we perfom 
the superspace integration, which amounts to apply $D^2 \bar{D}^2$ 
to the entire expression and evaluate everything at 
$\theta = \bar{\theta} =0$, and keep only bosonic terms. 

The covariant term in eq. (\ref{BGJ3}) gives 
\beq
L ~\rightarrow~ \frac{i}{8\pi^2} \int d^4x
 \, {\rm Tr} ( \, \l \, f^{\a \b} \star f_{\a \b} \,)
\label{covbosonic}
\eeq
In the reduction of the consistent part only terms at the
most linear in $V$ contribute. 
Therefore, expanding up to the linear order, we are led to compute
the expression
\bea 
&& \frac{i}{8\pi^2} \, \int \, d^4x \,
D^2 \, \bar{D}^2 \Big\{ \,
\frac{1}{3} \, \int_0^1 \, dy \, y^2 \, {\rm Tr}_s  \Big(
\, \L \star \pi^{\a} \star \omega_{\a} ~+~ (1-y) \,
(V \, \L) \star \pi^{\a} \star \omega_{\a} \nonumber \\
&&~~~~~~~~~~~~~~~+~ (1-y) \, \L \star \pi^{\a} \star (V \, \omega_{\a}) 
~-~ (1-y) \, V \star (\tilde{\pi}^{\dot \a}
\, \L) \star \tilde{\omega}_{\dot \a} \Big) \Big\} 
\label{linear}
\ena
After a long but straightforward calculation, from each term in (\ref{linear})
we have
\bea
&& \frac{i}{8\pi^2} \, \int \, d^4x \, \frac{1}{3} \, \int_0^1 \, dy \, y^2
\, D^2 \bar{D}^2 \, {\rm Tr}_s \, ( \L \star \pi^{\a} \star \omega_{\a})
\Big| \nonumber  \\
&& \rightarrow -\frac{i}{8\pi^2} \, {\rm Tr} \, \Big[ 
\, \frac{2}{3} \, \l \, f^{\a \b} \star f_{\a\b} - 
\frac{i}{12} \, \l \, \{A^{\a \dot \b} \, , \,  {A^{\b} }_{\dot \b}
\star f_{\a\b}\}_{\star} + \frac{i}{12} \, \l \, \{ f^{\a\b} \, , \, 
A^{~\dot \a}_{\b} \star A_{\a \dot\a} \}_{\star} \nonumber \\
&&~~~~~~~~ +~ \frac{1}{30} \, \l \, \{ A^{\a \dot \b} \, , \, 
A_{\b \dot\b} \star A^{\b \dot \a} \star A_{\a \dot\a} \}_{\star} ~+~ 
\frac{1}{60} \, \l \,
\{ A^{\b \dot\b} \star {A^{\a}}_{\dot \b} \, , \,  {A_{\b}}^{\dot \a} \star
A_{\a \dot \a} \}_{\star} \Big] \nonumber \\
&&~~~~~~~~~~~\nonumber \\
&& \frac{i}{8\pi^2} \, \int \, d^4x \, \frac{1}{3} \, \int_0^1 \, dy \, y^2
\, (1-y) \, D^2 \bar{D}^2 \, 
{\rm Tr}_s \, ( (V\L) \star \pi^{\a} \star \omega_{\a} ) \Big| \nonumber \\
&& \rightarrow ~ \frac{i}{8\pi^2} \, {\rm Tr} \Big[ \, \frac{i}{12}
\, \l \, \{ A^{\a \dot\b} \, , \,{ f_{\a}}^{\b} \}_{\star} \star A_{\b \dot \b}
~+~ \frac{1}{30} \, \l \, \{ A^{\a \dot\b} \, , \, A^{\b \dot\a} 
\star A_{\a \dot \a} \}_{\star} \star A_{\b \dot\b} \, \Big] \nonumber \\
&&~~~~~~~~~~\nonumber \\
&& \frac{i}{8\pi^2} \, \int \, d^4x \, \frac{1}{3} \, \int_0^1  \, dy \, y^2
\, (1-y) \, D^2 \bar{D}^2 \, 
{\rm Tr}_s \, ( \L \star \pi^{\a} \star (V \, \omega_{\a})) \Big| \nonumber \\
&& \rightarrow ~ -\frac{i}{8\pi^2} \, {\rm Tr} \Big[ \, \frac{i}{12}
\, \l \, \{ A^{\a \dot\b} \, , \,{ A^{\b}}_{\dot\b} \star f_{\a\b} \}_{\star}
~+~ \frac{1}{30} \, \l \, \{ A^{\a \dot\b} \, , \, {A^{\b}}_{\dot \b} 
\star {A_{\b}}^{\dot \a} \star A_{\a \dot \a} \}_{\star} \, \Big] \nonumber \\ 
&&~~~~~~~~~~~~~\nonumber \\
&& -\frac{i}{8\pi^2} \, \int \, d^4x \, \frac{1}{3} \, \int_0^1 \, dy \, y^2
\, (1-y) \, D^2 \bar{D}^2 \, 
{\rm Tr}_s \, ( V \star (\tilde{\pi}^{\dot \a} \, \L) \star 
\tilde{\omega}_{\dot \a}) \Big| \nonumber \\
&& \rightarrow ~ \frac{i}{8\pi^2} \, {\rm Tr} \Big[ \, \frac{i}{12}
\, \l \, \{ \bar{f}^{\dot \a \dot \b} \, , \, {A^{\b}}_{\dot \b} \}_{\star} 
\star A_{\b \dot\a} ~+~ \frac{1}{30} \, \l \, \{ A^{\a \dot\a} \star
{A_{\a}}^{\dot\b} \, , \, {A^{\b}}_{\dot\b} \}_{\star} \star  A_{\b \dot\a} 
\, \Big] 
\label{consbosonic}
\ena
The complete bosonic physical component is now obtained by adding
(\ref{covbosonic}), (\ref{consbosonic}) and their complex conjugates. Using
the identities \cite{super}
\bea
&& ~~~~~~~~~\epsilon^{abcd} ~=~ 
i [ C^{\a\d} C^{\b\g} C^{\dot\a \dot\b} C^{\dot\g
\dot\d} ~-~ C^{\a\b} C^{\g\d} C^{\dot\a \dot\d} C^{\dot\b \dot\g}]
\nonumber \\
&& F_{ab} ~=~ \Big[ C_{\dot\a \dot\b} f_{\a\b} ~+~ C_{\a\b} 
\bar{f}_{\dot\a \dot\b} \Big] \quad , \quad 
\Tilde{F}_{ab} ~=~ i \Big[ C_{\dot\a \dot\b} f_{\a\b} ~-~ C_{\a\b} 
\bar{f}_{\dot\a \dot\b} \Big]  
\ena
where $F_{ab} \equiv \pa_a A_b - \pa_b A_a - i [A_a , A_b ]_{\star}$ and 
$\Tilde{F}_{ab} \equiv \frac12 \e_{abcd} F^{cd}$, the total result 
is
\bea
{\cal A}_{\rm BGJ}^{bos}(\l) &=& \frac{1}{48\pi^2} \, \int d^4x \,
{\rm Tr} \Big\{ \, \l \, \Big[ F_{ab} \star \Tilde{F}^{ab} ~+~ 
i \Tilde{F}^{ab} \star A_a \star A_b \nonumber \\
&&~~ +~i A_a \star \Tilde{F}^{ab} \star A_b ~+~ i A_a \star A_b \star
\Tilde{F}^{ab} ~-~ \e^{abcd} A_a \star A_b \star A_c \star A_d \, \Big] \, 
\Big\} \nonumber \\
&&~~~~~~~~~~~~
\ena
As a consequence of the identity
\beq
\e^{abcd} \pa_a A_b ~=~ \frac12 \e^{abcd} F_{ab} ~+~ i \e^{abcd} A_a 
\star A_b 
\eeq
which follows from the definition of $F_{ab}$, 
the previous expression can be equivalently written as
\beq
{\cal A}_{\rm BGJ}^{bos}(\l) ~=~ \frac{1}{24\pi^2} \, \int d^4x \,
{\rm Tr} \Big\{ \l \, \e^{abcd} \pa_a \Big( \, A_b \star \pa_c A_d ~-~ 
\frac{i}{2} A_b \star A_c \star A_d \, \Big) \Big\}
\label{bosonican}
\eeq
This result coincides with the bosonic consistent anomaly derived in 
\cite{GBM}.


\newpage

\begin{thebibliography}{100}
\bibitem{NC}{C. Chu and F. Zamora, JHEP 0002 (2000) 022, hep--th/9912153; \\
S. Ferrara and M.A. Lled\'o, JHEP 0005:008,2000, hep--th/0002084; \\
S. Terashima, Phys. Lett. {\bf B482} (2000) 276, hep--th/0002119; \\
N. Grandi, R.L. Pakman and F.A. Schaposnik, {\em ``Supersymmetric 
Dirac--Born--Infeld theory in noncommutative space''}, hep--th/0004104; \\ 
D. Zanon, {\em ``Noncommutative Perturbation in Superspace''}, 
hep--th/0009196.}
\bibitem{HOOS}{ T. Hayashi, Y. Ohshima, K. Okuyama and H. Suzuki,
Prog. Theor. Phys. {\bf 100} (1998) 627, hep--th/9801062.}
\bibitem{OOSY}{ Y. Ohshima, K. Okuyama, H. Suzuki and H. Yasuta, 
Phys. Lett. {\bf B457} (1999) 291, hep--th/9904096.}
\bibitem{us}{S.J. Gates, jr., M.T. Grisaru, M.E. Knutt, S. Penati and 
H. Suzuki, {\em ``Supersymmetric Gauge Anomaly with General Homotopic 
Paths''}, hep--th/0009192.}
\bibitem{GBM}{J.M. Garcia--Bondia and C.P. Martin, Phys.Lett.
B479 (2000) 321-328, hep--th/0002171.} 
\bibitem{super}{S.J. Gates, Jr., M.T. Grisaru, M. Ro\v cek and W. Siegel,
{\em Superspace}, Benjamin Cummings, (1983) Reading, MA.}
\bibitem{us2}{S.J. Gates, jr., M.T. Grisaru and S. Penati, Phys. Lett.
{\bf B481} (2000) 397, hep-th/0002045.}
\bibitem{FGPS}{ S. Ferrara, L. Girardello, O. Piguet and R. Stora, 
Phys. Lett. {\bf B157} (1985) 179.}
 
\end{thebibliography}
\end{document}